\input harvmac

\def\Title#1#2{\rightline{#1}\ifx\answ\bigans\nopagenumbers\pageno0
\vskip0.5in
\else\pageno1\vskip.5in\fi \centerline{\titlefont #2}\vskip .3in}

\font\caps=cmcsc10

\noblackbox
\parskip=1.5mm

  
\def\npb#1#2#3{{\it Nucl. Phys.} {\bf B#1} (#2) #3 }
\def\plb#1#2#3{{\it Phys. Lett.} {\bf B#1} (#2) #3 }
\def\prd#1#2#3{{\it Phys. Rev. } {\bf D#1} (#2) #3 }
\def\prl#1#2#3{{\it Phys. Rev. Lett.} {\bf #1} (#2) #3 }
\def\mpla#1#2#3{{\it Mod. Phys. Lett.} {\bf A#1} (#2) #3 }
\def\ijmpa#1#2#3{{\it Int. J. Mod. Phys.} {\bf A#1} (#2) #3 }

\def\bb#1{{\tt hep-th/#1}}


           \def\CO{{\cal O}}


\def\dj{\hbox{d\kern-0.347em \vrule width 0.3em height 1.252ex depth
-1.21ex \kern 0.051em}}

\def\half{{1\over 2}\,}

\def\tr{{\rm tr\,}}

\def\ket{\rangle}
\def\bra{\langle}

\def\pt{\partial}
\def\li{\lambda_{\infty}}

\lref\rwitt{E. Witten, \npb {443}{1995}{85.}} 
\lref\rgreenf{M.B. Green, \plb {329}{1994}{435,} \bb{9403040.}}
\lref\rpolsem{J. Dai, R.G. Leigh and J. Polchinski, \mpla{4}
 {1989}{2073.}}
\lref\rpolrr{J. Polchinski, \prl{75}{1995}{4728,} \bb{9510017.}}
\lref\rgreent{M.B. Green, \plb {282}{1992}{380,} \bb{9210054.}}
\lref\rgreenb{M.B. Green, \plb {266}{1991}{325.}} 
\lref\raw{J. Atick and E. Witten, \npb{310}{1988}{291.}}
\lref\rpoint{E.F. Corrigan and D.B. Fairlie, \npb{91}{1975}{527\semi} 
M.B. Green, \plb{103}{1976}{333\semi}
I.R. Klebanov and L. Thorlacius, \plb {371}{1996}{51,}  \bb{9510200\semi}
J.L.F. Barb\'on, \plb {382}{1996}{60,} \bb{9601098.}}  
\lref\rbsus{T. Banks and L. Susskind, {\it ``Brane-Antibrane Forces,"} Rutgers
preprint  RU-95-87,  \bb{9511194.}}
\lref\rpolcom{J. Polchinski, \prd {50}{1994}{6041,} \bb{9407031.}}
\lref\rpolrev{J. Polchinski, S. Chaudhuri and C.V. Johnson,  {\it 
``Notes on D-Branes,"} Santa Barbara preprint NSF-ITP-96-003, 
\bb{9602052\semi}
J. Polchinski, {\it ``Tasi Lectures on D-Branes," } Santa Barbara 
preprint NSF-ITP-96-145, \bb{9611050\semi}  
E. Alvarez, J.L.F. Barb\'on and J. Borlaf, \npb{479}{1996}{218,}
 \bb{9603089.}}
\lref\rgreengut{M.B. Green and M. Gutperle, \npb {476}{1996}{484,}
 \bb{9604091.}}
\lref\rggut{M.B. Green and M. Gutperle, {\it ``Configurations of 
two D-instantons,"} Cambridge preprint DAMTP-96-110, \bb{9612127.}}
\lref\rggp{G.W. Gibbons, M.B. Green and M.J. Perry, \plb {370}{1996}{37,}
\bb{9511080.}}
\lref\rgreengas{M.B. Green, \plb {354}{1995}{271,} \bb{9504108.}}
\lref\rut{M.R. Douglas, D. Kabat, P. Pouliot and S. Shenker, 
\npb{480}{1997}{85,} \bb{9608024.}}
\lref\rfint{J.L.F. Barb\'on and M.A. V\'azquez-Mozo,  CERN-TH/96-361,
\bb{9701142.}}   
\lref\rbach{C. Bachas, \plb{374}{1996}{37,} \bb{9511043.}} 
\lref\rhs{G.T. Horowitz and A. Strominger, \npb{360}{1991}{197.}}
\lref\rshen{S. Shenker, {\it ``Another Lenght Scale in String Theory?"}
Rutgers preprint RU-95-53, \bb{9509132.}} 
\lref\rmalda{J.M. Maldacena, {\it ``Black Holes in String Theory",} 
Princeton Univ. Ph. D. Thesis, \bb{9607235.}}  
\lref\rbar{J.L.F. Barb\'on, {\it ``Fermion Exchange Between D-Instantons,"
} preprint CERN-TH/96-360, \bb{9701075.}} 
\lref\rdans{U.H. Danielsson, G. Ferretti and B. Sundborg, \ijmpa{11}{1996}
{5463,} \bb{9603081\semi}
D. Kabat and P. Pouliot, \prl {77}{1996}{1004,} \bb{9603127\semi}
G. Lifschytz, \plb {388}{1996}{720,} \bb{9604156.}} 
\lref\rbarton{B. Zwiebach, \plb {256}{1991}{22.}} 
\lref\rleigh{R.G. Leigh, \mpla {4}{1989}{2767.}} 
\lref\rdou{
M. Douglas and G. Moore, {\it ``D-Branes, Quivers, and ALE Instantons,"}
Rutgers preprint RU-96-15, \bb{9603167\semi}
M. Douglas, {\it ``Gauge Fields and D-Branes,"} Rutgers prep.  
RU-96-24 \bb{9604198.}}  

\line{\hfill CERN-TH/97-52}
\line{\hfill {\tt hep-th/9703138}}
\vskip 0.5cm

\Title{\vbox{\baselineskip 12pt\hbox{}
 }}
{\vbox {\centerline{Remarks on the Classical Size of D-Branes  }
}}

\centerline{$\quad$ {\caps J. L. F. Barb\'on
 }}
\smallskip

\centerline{{\sl Theory Division, CERN}}
\centerline{{\sl 1211 Geneva 23, Switzerland}}
\centerline{{\tt barbon@mail.cern.ch}}

 \vskip 1.0in

We discuss different criteria for ``classical size" of extremal
Dirichlet $p$-branes in type-II supergravity. Using strong--weak
 coupling
duality, we find that the size of the strong-coupling region 
at the core of the $(p<3)$--branes, is always given
by the asymptotic string-scale, if measured in the weakly 
coupled dual string-metric. We also point out how the 
eleven-dimensional Planck scale arises in the classical
 0-brane solution,
as well as the ten-dimensional Planck scale in the D-instanton
solution. 
\Date{March 1997}  



One of the interesting properties of Dirichlet-branes (D-branes) 
\refs\rpolsem, \refs\rpolrev,  is
their sharp character in space-time. Since they are perturbatively
defined as a boundary condition for open-string propagation, they
have no ``bare" size. Still, they interact
at a perturbative level 
via virtual-string exchange, and there is an effective string-scale
``halo" which represents a minimun size for perturbative string
probes (with the notable exception of the D-instanton \refs\rpoint).        
A minimun distance of the order of the string scale is also singled
out when considering the short-distance interactions of branes and
anti-branes \refs\rgreenb, \refs\rgreenf, \rbsus, \refs\rbar, due
to a local analogue of the well-known Hagedorn phenomenon (see for
example \refs\raw). 

On the other hand, if we use D-branes themselves as probes 
\refs\rbach, \refs\rdou, \refs\rdans,
\refs\rut, we can isolate sub-stringy scales in the 
slow scattering of D-branes. Such sub-string scales were
predicted on general grounds in \refs\rshen.

In this note we collect some heuristic arguments based on the
analysis of low-energy supergravity solutions. We will consider the
relevant dynamical scales as being 
 associated with the strong-coupling dynamics
 measured by the local string coupling $\lambda (r) = e^{\phi(r)}$,
where $\phi(r)$ is the dilaton expectation value in the configuration
of interest. Using string dualities locally, we can always turn a
region of extreme strong coupling into a weakly coupled region by
an appropriate change of variables. Therefore, we will be concerned
with the points of ``trully strong coupling", namely those  with a
local string coupling of order unity $\lambda (r) 
\sim \CO (1)$. In this way, we include non-perturbative effects
through the non-linearities of the low-energy supergravity theory.         

At first sight, these 
 considerations have at best a heuristic value, because the
low-energy supergravity solution should not be trusted at short distances.
However, we should not forget that the supergravity solutions are protected
by supersymmetry. Moreover, because of the strong-weak coupling dualities
of the ten-dimensional supergravities, we never have to consider a local
region of space-time at extreme strong coupling. For these reasons,
 we think
that the classical analysis, if properly interpreted, is capable of
capturing the essential scales, at least in order of magnitude.

The relevant bosonic terms in the string-frame supergravity 
  action are (excluding the $p=3$ case):   
\eqn\action
{S = \int \sqrt{|g|} e^{-2\phi} \left( R - 4(\partial \phi)^2 \right) +
\int \sqrt{|g|} { e^{2 a \phi} \over 2 (8-p)!} \, F^2_{8-p} .       }
Here the parameter $a=0$ corresponds to R--R backgrounds in 
type-II, and $a = -1$ corresponds to the heterotic low-energy theory.
The R--R solutions, in magnetic form, 
 with a vanishing asymptotic dilaton, are given by \refs\rhs, \refs\rggp:     
\eqn\sol
{\eqalign{ ds^2 =& f^{-1/2} dx^2_{L} + f^{1/2} \left( dr^2 + r^2
d\Omega^2_{8-p} \right) \cr
e^{-2\phi'} =& f^{p-3 \over 2} \cr
F'_{8-p} =& C_p \,d\Omega_{8-p}\cr
f =& 1 + {1\over 7-p} {C_p \over r^{7-p}},  }}  
with $dx_L$ the coordinate elements in the world-volume directions, and
$r$ the transverse radial coordinate. The constant $C_p$ is proportional
to the charge $Q$, or total number of coincident branes.  
We can switch on an asymptotic dilaton and coupling constant
$\lambda_{\infty} = e^{\phi_{\infty}}$  by the substitutions $\phi
= \phi' + \phi_{\infty}$ and $  F' = \lambda_{\infty}^{a +1} F$.
The action becomes
$$
S = {1\over \li^2} S({\rm primed} \,\,{\rm fields}) \sim {C_p \over \li^2}  
.$$ 
Therefore, for R--R branes 
  we need to take $C_p \sim \li Q$, in order to have the
correct  R--R scaling of the action 
$S\sim Q/ \li$.

 The same result follows from  the 
ADM definition of tension (see for example ref. \refs\rmalda).   
 Consider a completely wrapped R--R $p$-brane,  
which looks like a black hole in $d=10-p$ dimensions. The corresponding
string-frame mass can be calculated using T and U-duality:
\eqn\ma{ M_p =  {Q R^p \over \li\,( \alpha')^{p+1 \over 2}}, }
 for a symmetric torus of
 radius
$R$. This is the same mass that one finds in the so-called 
``modified Einstein frame", which decouples the dilaton, but coincides
asymptotically with the string frame 
\eqn\mef{ (g_{\mu\nu})_{\rm str}
 = \, e^{4\phi' \over d-2} \, (g_{\mu\nu})_{ME}.   } 
Using the   solutions above, 
 we have the following ADM matching at large
distances:    
\eqn\adm{-(g_{00})_{ME}
 = f^{3-d \over d-2} \rightarrow 1-{1\over d-2}
\,{C_p \over r^{d-3}} = 1- {1\over (d-2)\,|S^{d-2}|}\,
{16\pi G_d M_p \over  r^{d-3}},}
with $|S^{d-2}|$ the volume of the unit $(d-2)$-sphere and  $G_d$ the
$d$-dimensional Newton constant,  given by the standard Kaluza--Klein
reduction, $G_d = G_{10} / (2\pi R)^p $, from the ten-dimensional Newton
constant. The latter is fixed by string-fivebrane  Dirac duality to be
 $G_{10} =8\pi^6 \alpha'^4 \li^2$.
With this information, it is easy to solve for $C_p$ with the result
\eqn\cp{C_p = \li Q \, {(2\pi\sqrt{\alpha'})^{7-p} \over |S^{8-p}|}.}

The upshot of the preceding discussion is simply that  we must
 use the classical metric in  \sol\ with $C_p \sim Q \li$ and a 
local string coupling  
\eqn\stc{
\lambda (r) = \li \, f(r)^{3-p \over 4}
.}

For $p< 3$,  the string
metric at the core 
 exhibits a neck followed by an infinite non-compact space
 extending an infinite proper distance, i.e.  $\int dr \sqrt{g_{rr}}$ 
diverges as we approach the origin. The volume of the transverse
sphere $S^{8-p}$ at the point $r$ is given by
$$
V_{\Omega} (r) = \left(\sqrt{r^2 \,f^{1/2} (r)}\right)^{8-p}
\cdot |S^{8-p}|, $$
and scales as $r^{8-p}$ for large $r$, and as $r^{(p-5)(8-p)}$ close
to the origin. Therefore, there is a minimun transverse volume 
 of the neck, scaling as  
$
V_{\Omega}^{\rm min} \sim (Q \li)^{8-p \over 7-p} 
$, 
thereby defining an effective proper radius\foot{Unless otherwise
specified, lengths are measured in units of the string length $\sqrt{
\alpha'}$.}   
 \eqn\trad{ R_{\min}
 \sim (Q\li)^{1\over
7-p}.} 
   At this point $f_{\rm min} \sim \li^0 \sim \CO (1)$,    
 and  the local string
coupling is of the same order as the asymptotic coupling. This
means that, 
in some  sense,  we can assign a small size to the D-brane soliton,
within the realm of perturbative string theory.
The more interesting scales, however, are associated to the region  
of local strong coupling.

For the self-dual 3-brane, the dilaton is uniform and asymptotic weak
coupling ensures weak coupling throughout space. For $p>3$, the
situation is even better, since the horizon region is always weakly
coupled. On the other hand, for $p<3$ we have 
strong coupling at the origin. Thus, the
 other natural scale we can define is determined by the  
 point $r_c$ at which the coupling becomes strong; $\lambda (r_c)
 \sim \CO (1)$. It 
scales  as  
\eqn\gs{
r_c \sim \li^{1\over 3-p}
\,Q^{1\over 7-p}.}   
An analogous scale is defined by requiring the effective open-string
coupling $\lambda_o \sim \lambda (r) Q$ to be of order unity:
\eqn\gso{ r_o \sim (Q\li )^{1\over 3-p} .} 

In view of the results of ref. \refs\rut, it
 is very suggestive that \gs\ and \gso, defined as the radial coordinate
where strong-coupling effects take place, gives the
eleven-dimensional Planck length for the case of the 0-branes; 
$\ell_{11} \sim \lambda_{\infty}^{1/3} \sqrt{\alpha'}$. Other interesting
cases are the D-instanton ($p=-1$), for which the strong-coupling
radial coordinate is of the order of the ten-dimensional Planck
scale\foot{This agrees with the dynamical analysis of \refs\rggut.},  
 $\ell_{10} \sim \li^{1/4} \sqrt{\alpha'}$, and the D-string
 ($p=1$),
 whose ``size" under this criterium scales as $\li^{1/2} \sqrt{
\alpha'}$, again in agreement with the dynamical analysis of ref. \rut. 
        
We regard these agreements as significative, even if 
  \gs\ and \gso\ do not define a proper distance in any of the 
non-linearly corrected string metrics for $p<3$. At the end of
this note we will comment on the significance of the
scales defined by $r_c$ and $r_o$.
 However, at this stage, it is more natural to
perform a strong-weak coupling duality transformation to rewrite
the geometry of the core region 
 in terms of weakly coupled variables. Since such 
duality transformations involve a Weyl rescaling of the string
metric, the infinite non-compact space at the core is replaced
by a finite patch with a 
 pointlike singularity, which can be matched to a delta-function
source. It is then natural to define a strong-coupling radius
as the proper distance    
 from the origin to the self-dual point
where the local string coupling is of order unity, i.e. the
point of ``trully strong coupling", measured in the weakly coupled
string metric.

In the case of $p=-1, 1$ branes, we have a type II-B theory, and the
dual  metric is found by the usual requirement that the
Einstein metric is invariant, where, in ten dimensions $
e^{-\phi/2} g_{\mu\nu} = g^E_{\mu\nu} = e^{-{\tilde \phi}/2} {\tilde
g}_{\mu\nu} $, and $\phi = -{\tilde \phi}$. There is an interesting
 cancellation
of all string coupling powers, which results in a proper distance
proportional to the D-brane charge;   
\eqn\dosa{\eqalign{
{\tilde L} = & \int_0^{r_c} dr \sqrt{{\tilde g}_{rr}} = \int_0^{r_c}dr  
\lambda (r)^{-\half}\,\sqrt{g_{rr}}  
\sim \li^{-\half} \int_0^{r_c} dr f(r)^{p-1 \over 8}\sim   \cr    
\sim &\, \li^{-{1\over 2}+ {p-1 \over 8}}
\,\,Q^{p-1 \over 8}\,\, r_c^{{(7-p)(1-p) \over 8} + 1} \sim Q \li^0
\sim Q.}}  

On the other hand, when considering $p=0,2$ branes, we have a type II-A
theory, and the strong-coupling limit is governed by eleven-dimensional
supergravity.
 The mapping between the II-A string metric and the ten-dimensional
 part
of the ${\rm Sugra}_{11}$ metric is \refs\rwitt  
$$
{\tilde g}_{rr}  = e^{-{2\phi \over 3}} g_{rr},  
$$
with $\phi$ the II-A dilaton. Remarkably, the powers of
$\lambda_{\infty}$ still conspire to cancel the coupling dependence:
\eqn\dosb{\eqalign{ 
{\tilde L} = & \int_0^{r_c} dr \sqrt{{\tilde g}_{rr}} = \int_0^{r_c}
dr \lambda (r)^{-{1\over 3}} \sqrt{g_{rr}} = \li^{-{1\over
3}} \int_0^{r_c} dr f(r)^{p\over 12} \sim      \cr  
\sim  & \, \li^{-{1\over 3} +
{p\over 12}}\,\, Q^{p\over 12}
\,
\, r_c^{{p(p-7) \over 12} +1} \sim Q \li^0
\sim Q. }}  
Thus, the so-defined strong-coupling radius of R--R $ p$-branes
 is always
${\tilde L}_{RR} \sim \sqrt{\alpha'} = \ell_{\rm str}$,  
where $\ell_{\rm str}$ is the string scale set by the asymptotic
weakly coupled string theory.
 We find it remarkable that this behaviour also extends to the
type-IIA case, where the dual, weakly coupled metric, is {\it not}
a string sigma-model metric. 

 This situation also  
applies to  the case of the heterotic
5-brane, whose metric is 
$$
ds^2 = dx^2_{L} +\left(1+{r_+^2 \over r^2} \right)\left( dr^2 + r^2
d\Omega_3^2 \right)
,$$
and the local string coupling, 
 $\lambda (r) = \li \sqrt{1+ r_+^2 / r^2 }$. In
this case we have to choose $r_+ \sim \li^0$, because the action must
have an NS scaling $S\sim \li^{-2}$.
 Then, the strong-coupling $\lambda (r_c) \sim \CO (1)$
point lies at $r_c \sim \li$, and the proper distance in the dual
type-I metric \refs\rwitt  
$$
{\tilde L}_{NS} = \int_0^{r_c} dr \lambda (r)^{-\half}\,
 \sqrt{g_{rr} } \sim
\li^{-1} 
\int_0^{r_c} dr \sqrt{\lambda (r)} \sim  \sqrt{r_c \over \li  } \sim \li^0 
\sim \CO (1).$$
It is interesting to notice that the same results follow in the
Einstein frame:
$$
L_E = \int_0^{r_c} dr \sqrt{g^E_{rr}} = \int_0^{r_c} dr 
 \lambda(r)^{-{1\over 4}}
\sqrt{g_{rr}} 
.$$
One obtains $L_E \sim \sqrt{\alpha'}$ in both cases.
Of course, this is just the Planck length in Einstein units. We can
pass to the modified Einstein frame, which rescales the Einstein-frame
lengths by the constant factor $\li^{1/4}$, and find the standard
form of the Planck length in string units. 
Being equivalent to the string frame at large distances, 
it is not completely trivial that the modified Einstein frame should
define a Planckian strong-coupling radius.   

The short-distance scales defined by \gs\ and \gso\ 
 correspond to proper
distances in the flat metric that approaches the string metric
at radial infinity. This is interesting because, in the D-brane
picture, we have a flat Minkowski space-time all the way down
to the D-brane core, where the Dirichlet boundary condition is
enforced. In other words, the D-brane is really a source for
closed string massless fields, and the back-reaction is not taken
into account in the definition of boundary states by ``bare" 
Dirichlet conditions on the world-sheet fields. 

Dynamically, the flat metric is appropriate for describing the
slow scattering of test D-branes (see \refs\rut, \refs\rdans). To
some extent, this is  
 true even after we take into account the non-linearities. The
action for a test D-brane in the  non-linear
 background of another D-brane can be approximated by simply 
substituting the fields of   \sol\ into the Born-Infeld action   
\refs\rleigh. 
For a rigid, straight,  
D-brane in a physical gauge, i.e. $x^0 =t$, ${\vec x}_L
= {\vec \sigma}$, ${\vec x}_T = {\vec x}_T (\tau)$, the factors
of the profile function, $f(r)$, cancel out  such that one finds
\eqn\bi{\eqalign{ S_{BI} =& -C\int d\tau d^p {\vec\sigma}\,
\, e^{-\phi}\,\sqrt{-{\rm det}(g_{\mu\nu} 
\pt_{\alpha} x^{\mu} \pt_{\beta} x^{\nu} )} \cr
 =&  -{C V_p \over \li} \int
d\tau \,{1\over f} + {C V_p \over \li} \int d\tau \,
 \half \delta_{ij} {\dot x}_T^i {
\dot x}_T^j  + 
\CO ({\dot {\vec x}}_T^4 ).}}       
The first term is the action of the D-brane at rest, whereas the
 second term is simply the non-relativistic action for a
slow-moving object with a mass $M= C V_p /\li$, in a
space with flat transverse metric. 
 
From the point of view of the closed-string sector, the D-brane
boundary condition does not define an exact background, due to
the non-vanishing tadpoles,  
\eqn\tad
{\bra V_{\rm massless} | B \ket \neq 0
,}
of the massless closed-string states in the presence of the
D-brane boundary state.     This means that the series of
string diagrams,   
 with Dirichlet boundary conditions and free strings propagating
in Minkowski space-time,  is really an expansion
around an approximate solution of the closed-string equations.   
To see how this explains some features of the Dirichlet-diagram
expansion, consider for definiteness the case of a single D-instanton.
The ``tree" effective action is given by \refs\rpolcom, \refs\rgreengas, 
\refs\rfint:  
\eqn\tr{
\Gamma_{\rm tree} = \sum_{N=1}^{\infty} {1\over N!} W_N, }
where $W_N$ is a sphere amplitude 
 with $N$ Dirichlet boundaries in the flat
conformal field theory. If we perform a string field theory 
(SFT)  decomposition
as in \refs\rbarton,        
 we can write (see also \rut):  
\eqn\fac
{\Gamma_{\rm tree} = \sum_N \sum_{\{\alpha\}} {1\over N!} W_{\alpha_1 
\cdots \alpha_N} \, \prod_{i=1}^{N} {1\over -\pt^2 + M_{\alpha_i}^2} \,
\bra V^*_{\alpha_i} | B\ket, }
where $W_{\{\alpha\}}$ denotes the generalized SFT irreducible vertex
  with the indices $\alpha_i$ labeling  
 states defined in the flat conformal field theory\foot{Notice
 that we are inserting the full tower of
massive closed-string states, and not only the massless ones, as would
be appropriate at large distances.}, and $\bra V^*_{\alpha_i} |B\ket$ 
is an off-shell SFT generalization of the perturbative tadpoles in 
\tad.
 Using now the fact that correlators of vertex operators
give directly the amputated string field theory amplitudes, we can
write \fac\ in the form
\eqn\ottt{
\Gamma_{\rm tree} = \sum_{\{\alpha\}} {1\over N!} W_{\alpha_1 \cdots
\alpha_N} \delta \Psi_{\alpha_1}^{c\ell} \cdots \delta \Psi_{\alpha_N}^{
c\ell}.}   
Here $$\delta \Psi_{\alpha}^{c\ell} \equiv \Psi_{\alpha}^{c\ell}
 ({\rm inst}) - \Psi_{\alpha}^{c\ell}    
({\rm vac})$$ is the difference between the corresponding
string field at the flat Minkowski CFT, and the exact closed-string
instanton background. Therefore, if we interpret the flat, off-shell  
SFT vertices  as functional derivatives,   
$$
W_{\alpha_1 \cdots \alpha_N} \sim {\delta^{(N)} \Gamma_{\rm tree}
 \over \delta \Psi_{
\alpha_1} \cdots \delta \Psi_{\alpha_N}}, 
$$
we find the expansion of the effective action around an approximate
solution given by the flat CFT, with  sharp Dirichlet 
 boundary conditions at
the D-brane core.     
In particular, this explains why we can calculate the instanton
classical action by means of a diagrammatic computation (the disk
diagram being the leading term), a fact 
that has no analogue in field theory, unless we construct the diagrammatic
expansion around an approximate solution. Clearly, if the classical
action of the approximate solution vanishes, the entire classical action
of the exact instanton solution arises from tadpole contributions. 
 Scattering amplitudes can
be generated in the standard way, taking derivatives of the effective
action with respect to linearized deformations of the flat CFT.

We think that the above 
string field theory interpretation of the Dirichlet
diagrammatic expansion could shed some light on the dicotomy between
the conformal field theory descriptions of NS--NS solitons, as 
oppossed to the D-brane construction for R--R solitons. Indeed, the
previous construction treats the R--R instantons in a more familiar
way, from the field theoretical point of view.    

Comming back to the issue of strong-coupling length scales, we
conclude that the appropriate notion of ``classical size" of 
D-branes depends on the probe we use to measure geometry. If we
consider closed string propagation in the background of a D-brane,
non-linear effects induce a strong-coupling scale that, when 
measured in the weakly coupled metric, still corresponds to the
string length $\sqrt{\alpha'}$, the size of the perturbative ``halo". 
On the other hand, when considering the slow scattering of D-branes
as a probe of geometry, the effective transverse metric is the
flat metric \bi. Thus, we can interpret the sub-stringy scales of
\refs\rut\ as the strong-coupling scales \gs\ and \gso, determined
 directly from the classical dilaton profile in eq. \stc.

\newsec{Acknowledgements} 
It is a pleasure to thank  
 D.J. Gross and I.R. Klebanov for useful discussions 
at the early stages of this work.

\listrefs
\bye